# Strong luminescence quantum efficiency enhancement near prolate metal nanoparticles: dipolar versus higher-order modes

H. Mertens and A. Polman[*]

Center for Nanophotonics, FOM Institute for Atomic and Molecular Physics,

Kruislaan 407, 1098 SJ Amsterdam, The Netherlands

**Abstract**

We present a theoretical study of the radiative and nonradiative decay rates of an optical emitter in close proximity to a prolate-shaped metal nanoparticle. We use the model developed by Gersten and Nitzan, that we correct for radiative reaction and dynamic depolarization and extend for prolate particle shapes. We show that prolate-shaped metal nanoparticles can lead to much higher quantum efficiency enhancements than corresponding spherical nanoparticles. For properly engineered emitter-nanoparticle geometries, quantum efficiency enhancements from an initial value of 1% (in absence of the nanoparticle) to 70% are feasible. We describe the anisotropy-induced enhancement effects in terms of large field enhancements near the metal tips that cause strong coupling to the (radiative) dipolar modes. For increasing particle anisotropy, a strong spectral separation between radiative dipolar and dark higher-order modes occurs, which leads to higher radiative efficiencies for anisotropic particles. In addition, we demonstrate that for large (> 100 nm) nanoparticles, the influence of Ohmic losses on plasmon-enhanced luminescence is substantially reduced, which implies that, if prolate shaped, even lossy metals such as Al and Cu are suitable materials for optical nano-antennas.

---

[*] Electronic address: polman@amolf.nl

# I. Introduction

The control of spontaneous emission by metal nanostructures is a subject of intense research.[1,2,3,4,5] The interest in this phenomenon is largely motivated by the fact that the interaction between optical emitters and plasmon modes, i.e., collective conduction electron oscillations in a metal, can enhance the emission rate,[3,5] polarization state[6,7] and directionality[8] of the emitted light. Metal nanostructures are therefore powerful tools for tailoring the performance of solid-state light sources.

With the ongoing interest in the electromagnetic interaction between emitters and metal nanostructures, there is a growing demand for theoretical studies that provide insight in the physical factors that determine the performance of nano-antenna geometries.[9,10,11,12,13,14] In a previous paper,[15] we have shown that the model developed by Gersten and Nitzan (GN)[16,17,18] for small metal nanoparticles can also accurately describe decay rate enhancements near larger (several 100-nm diameter) metal nanospheres, provided that the well-known correction factor for radiative reaction and dynamic depolarization is implemented.[19,20,21] A key advantage of the GN model is that it can be generalized to spheroidally shaped particles, using expansions in terms of an orthogonal set of eigenfunctions. Such an expansion is not known for exact electrodynamical theory.[22] Consequently, the improved GN model, i.e. including the corrections for radiative reaction and dynamic depolarization, is a unique analytical method to investigate plasmon-enhanced luminescence near spheroidal metal nanoparticles.

In this paper, we apply the improved Gersten and Nitzan model to compare the decay rates of an optical emitter near spherical and prolate metal nanoparticles.[23] Based on these calculations we conclude that prolate metal nanoparticles are more suitable for resonantly enhancing the quantum efficiency of a low-quantum-efficiency emitter than spherical metal nanoparticles. We explain this effect in terms of the size and shape dependencies of the electromagnetic properties of the nanoparticle plasmon modes. In addition, we show that for prolate metal nanoparticles, the influence of Ohmic losses is strongly reduced compared to the spherical case. Therefore, also metals that are more lossy than Ag and Au, such as Al and Cu, can be suitable materials for optical nano-antennas. While this paper focuses on prolate nanoparticles, the main conclusions are valid for anisotropic shapes in general.

The paper is organized as follows. In Sec. II we summarize the concepts behind the improved Gersten and Nitzan model. In Sec. III, we compare the quantum efficiency enhancement of a low-quantum-efficiency emitter near a prolate Ag nanoparticle to that of a similar emitter near a spherical Ag nanoparticle. We observe that the quantum efficiency enhancement for the prolate geometry exhibits two features that deviate from the result for the sphere geometry. These two features are analyzed in detail in the subsequent two sections. In Sec. IV, we focus on the nanoparticle size and shape dependence of the interaction between the emitter and the dipole plasmon mode. In Sec. V, we discuss the shape dependence of the interaction between the emitter and higher-order plasmon modes. We investigate how energy transfer to these lossy plasmon modes can be minimized. Finally, conclusions are presented in Sec. VI.

## II. Method

The modifications of the radiative decay rate $\Gamma_\text{R}$ and the total decay rate $\Gamma_\text{TOT}$ of a dipole emitter in close proximity to either a spherical or a prolate metal nanoparticle can be calculated with the improved Gersten and Nitzan model. In this model, the decay rate modifications are calculated based on a two-step approach. In the first step, the electromagnetic interaction between source dipole and nanoparticle is analyzed based on the quasistatic approximation.[24] In this approximation, all dimensions are assumed to be much smaller than the wavelength of light, so that retardation effects can be accounted for by simply modifying the quasistatic polarizability of the nanoparticle with a correction factor for radiative reaction and dynamic depolarization.[19,20] This correction factor has previously been applied for improving the quasistatic description of scattering. In the second step, an effective dipole moment of the coupled system is identified. This effective dipole moment, which comprises a vectorial superposition of the source dipole moment and the induced dipole moment, is used to calculate the radiated power. By normalization to the power radiated by an uncoupled source with identical dipole moment, the modification of the radiative decay rate is obtained.[25] The total decay rate, which includes both radiative decay and nonradiative decay that is associated with dissipation in the metal, is determined by calculating the power that the source dipole transfers to the various plasmon modes of the metal nanoparticle and to free-space radiation. An overview of the mathematical expressions for the decay rate modifications for an emitter in the vicinity of a metal sphere, according to the improved Gersten and Nitzan model, is given in the appendix. The corresponding expressions for spheroidal nanoparticles can be obtained by applying a similar correction[26] to the original GN expressions for spheroidal nanoparticles.[16,17,18,23] Limitations of the improved GN model are (1) the fact that multipole radiation is not described, and (2) the fact that interference between source dipole and induced dipole is neglected. The latter limitation is discussed in more detail in Sec. IV.

In addition to the results based on the improved GN model, we also show results for spherical particles that are based on exact electrodynamical theory.[27,28] These results serve as benchmarks for the improved Gersten and Nitzan model. An extensive discussion on this type of calculations can be found in Ref. 15.

### III. Quantum efficiency enhancement near a Ag nanoparticle

In this section, we consider the modifications of the quantum efficiency $\eta = \Gamma_R / \Gamma_{TOT}$ of an emitter in the vicinity of either a spherical or a prolate Ag nanoparticle. The emitter and the nanoparticle are both embedded in a homogeneous dielectric with a refractive index of 1.5 to reflect the often used experimental case of silica glass. We assume the emitter has an internal luminescence quantum efficiency of 1% in absence of the nanoparticle. This relatively low quantum efficiency was chosen to illustrate some important aspects of plasmon-enhanced luminescence.

Figure 1(a) shows the quantum efficiency of the emitter near a 60-nm-diameter Ag sphere as function of emitter position in a two-dimensional cross section through the particle's major axis. The quantum efficiency was calculated with the improved GN model, by taking into account coupling to all multipole modes up to $l = 100$. The orientation of the dipole was taken parallel to the major axis (z), and the emission wavelength of 500 nm was chosen such that the emission is resonant with the dipole plasmon mode of the Ag sphere. The optical data for Ag were taken from Ref. 29. Figure 1(b) shows a similar plot as Fig. 1(a), but for a prolate Ag nanoparticle with an aspect ratio (i.e. major axis length divided by minor axis length) of 2.5, with a volume equal to that of a 60-nm-diameter sphere. The emission wavelength was set to 794 nm, which is resonant with the longitudinal dipole plasmon mode. This resonance is shifted towards the infrared compared to the resonance of the Ag sphere, due to shape anisotropy.[30]

Figs. 1(a) and (b) both show that the quantum efficiency is substantially enhanced close to the nanoparticle, in particular for emitter positions for which the source dipole has a significant component perpendicular to the adjacent metal surface. Moreover, the quantum efficiency enhancement is much larger close to the sharp tips of the prolate than close to the sphere, despite the fact that both geometries are evaluated at the optimum wavelength. To make this comparison more quantitative, Figure 1(c) shows line traces of the quantum efficiency that are taken along the dashed lines plotted in Figs. 1(a) and (b). The solid lines in Fig. 1(c) represent calculations in which coupling to all multipole modes up to $l = 100$ are taken into account; the dashed lines represent calculations in which only coupling to the dipole mode is considered. For the sphere geometry, the quantum efficiency is enhanced from 1% to 27% for an emitter-nanoparticle separation of about 5 nm. As can be seen, this enhancement is essentially due to coupling between the emitter and the dipole plasmon mode. For emitter-nanoparticle separations smaller than 5 nm, the quantum efficiency drops due to coupling between the emitter and dark higher-order plasmon modes, as discussed in detail in our previous paper. For the prolate geometry the quantum efficiency is enhanced from 1% to 65%. This significant improvement compared to the spherical case is due to two effects. First of all, the coupling between the emitter and the dipole plasmon mode is more effective for the prolate than for the sphere geometry, as can be seen from the dashed lines. This effect is due to the shape-induced field enhancement near the sharp tip of the prolate, as is explained in Sec. IV. Second, the drop in quantum efficiency close to nanoparticle surface sets in at smaller separations for the prolate than for the sphere geometry. For example, the distance to the metal surface at which the quantum efficiency reaches half the maximum value is 2.2 nm for the sphere, and only 0.8 nm for the prolate. Quenching due to coupling with higher-order plasmon modes is thus less effective for prolate nanoparticles. The reduced coupling to higher-order plasmon modes is analyzed in detail in Sec. V.

## IV. Improved coupling between emitter and dipole plasmon mode

Figure 2(a) shows the wavelength of maximum radiative decay rate enhancement associated with coupling to the longitudinal dipole plasmon mode of a prolate Ag nanoparticle versus equivalent nanoparticle diameter, for four different aspect ratios. The term *equivalent diameter* refers to the diameter of a sphere with identical volume, and is thus a measure for the nanoparticle size. The lines with symbols shown in Fig. 2(a) are calculated with the improved GN model; the solid line for aspect ratio 1 is obtained from exact electrodynamical theory. The good agreement between both results for aspect ratio 1 indicates the applicability of the improved GN model to the nanoparticle sizes under consideration. Fig. 2(a) shows a redshift of the wavelength of maximum radiative decay rate enhancement with increasing diameter for all four aspect ratios. This redshift is directly related to the dipole resonance redshift for increasing size, which is mainly attributed to retardation of the depolarization field across the nanoparticle. In addition, the wavelength of maximum radiative decay rate is further redshifted for increasing aspect ratio, as was already discussed in the context of Fig. 1. We note that the wavelength of maximum radiative decay rate enhancement is nearly independent of emitter-sphere separation for the sub-wavelength distances that we are interested in.

Figure 2(b) shows the maximum quantum efficiency calculated at the wavelength shown in Fig. 2(a), by taking into account coupling to all plasmon modes up to $l = 100$. The maximum quantum efficiency refers to the maximum values in quantum-efficiency-versus-distance plots, as shown in Fig. 1(c), and thus represents different (optimized) emitter-sphere separations, ranging from 3 to 10 nm. The data were calculated for an emitter that has a quantum efficiency of 1% in the absence of the sphere, and for a dipole orientation that is parallel to the major axis. Figure 2(b) illustrates that for the sphere geometry there is an optimal diameter of about 55 nm at which the quantum efficiency is enhanced most: from 1% to 27%. This size dependence of the quantum efficiency enhancement for the sphere geometry has been discussed in detail in our previous paper. For increasing aspect ratio, we see that the maximum quantum efficiency reaches higher values, up to 70%. Prolate metal nanoparticles are thus more effective for enhancing the quantum efficiency of a low-quantum-efficiency emitter than spherical nanoparticles. The effect that the quantum efficiency of an emitter coupled to anisotropic metal nanoparticles can be larger than 80% has been derived from FDTD simulations by Rogobete *et al.* (for a reference quantum efficiency in absence of nanoparticles of 100% in that case). In the remainder of this section we explain, based on the analytical GN method, why the coupling between emitters and prolate metal nanoparticles can result in such high quantum efficiencies.

Figure 3 shows the size and shape dependencies of the two processes that determine the quantum efficiency enhancement at the optimum emitter-nanoparticle separation. Fig.

3(a) displays the excited state decay rate associated with coupling to the dipole plasmon mode $\Gamma_{TOT,DIP}$ normalized to the radiative decay rate in absence of the metal nanoparticle $\Gamma_R^{REF}$, plotted as a function of equivalent diameter for four different aspect ratios (see the Appendix for an explanation on how this dipole contribution is determined). This decay rate is calculated for the optimized distance, which is between 3 and 10 nm, and for the optimum wavelength shown in Fig. 2(a). For the sphere geometry, the decay rate increases in magnitude for decreasing equivalent diameter. This trend indicates that resonant coupling between the emitter and the dipole plasmon mode is stronger for smaller spheres, as discussed in detail in Ref. 15. The same trend can be observed for prolate nanoparticles, for which the absolute values of the decay rate are one to two orders of magnitude higher. This is an important advantage of prolate (and in general: anisotropic) nanoparticles, as will be discussed below.

Figure 3(b) shows the fraction of the total power coupled to the dipole mode that is reradiated $F$, defined as $F = \Gamma_{RAD,DIP} / \Gamma_{TOT,DIP}$. According to the exact electrodynamical theory for spheres (solid line), $F$ is smaller than 10% for sphere diameters below about 20 nm, and reaches values higher than 90% for sphere diameters larger than 85 nm. The latter means that radiation damping dominates the decay of the dipole plasmon mode for sufficiently large Ag nanoparticles. $F$ for the sphere geometry as calculated with the improved GN model (top graph in fig. 3(b)) deviates from the exact result and reaches values up to 1.3. The fact that values higher than 1 are reached is a deficiency of the improved GN model, which is caused by the fact the radiative rate of the emitter is determined by approximating the system of emitter and nanoparticle as an effective dipole. As a consequence, interference effects between the source dipole and the dipole induced in the nanoparticle are not taken into account. The lines for prolate nanoparticles do not reach values as high as the result for the spherical nanoparticle. This difference is attributed to the fact that for prolate nanoparticles, the induced dipole is much larger than the source dipole, due to improved emitter-plasmon coupling. As a consequence interference is less important, and the improved GN model is thus more accurate for prolate nanoparticles. Despite the deficiency of the model, we can conclude from Fig. 3(b) that radiation damping dominates the decay of the dipole plasmon mode of sufficiently large prolate nanoparticles. This conclusion is an important guideline for the design of effective optical nano-antennas, as will be discussed next. Because we focus the rest of our analysis on prolate nanoparticles, the relatively large error in $F$ for spherical nanoparticles, as obtained from the improved GN model, does not affect our main conclusions.

Since quantum efficiency enhancement involves both coupling of the emitter to the plasmon mode, and outcoupling of plasmons to radiation, the opposite dependencies on equivalent diameter shown in Figs. 3(a) and (b) give rise to a trade-off. This explains the origin of the optimal diameter for quantum efficiency enhancement that is shown in Fig. 2(b). For prolate nanoparticles, the geometry-induced electromagnetic field enhancement near the sharp tip, often referred to as the lightning rod effect,[31] enhances the coupling between emitter and plasmon mode (Fig. 3(a)). As a consequence, also relatively large nanoparticles with their intrinsic high outcoupling efficiency of the dipole plasmon mode (Fig. 3(b)) can be strongly coupled to an emitter. This explains (1) why the maximum quantum efficiency shown in Fig. 2(b) increases and (2) why the corresponding optimal equivalent diameter shifts to larger values for increasing aspect ratio.

Figure 3(b) shows that for a nanoparticle with an equivalent diameter larger than about 50 nm the decay of the dipole plasmon mode is dominated by radiation damping instead of Ohmic losses. This suggests that also metals that exhibit higher Ohmic losses than Ag could be applied to enhance the quantum efficiency of a low-quantum efficiency emitter. In order to verify this hypothesis, Fig. 4 shows similar data as Fig. 2 now for Cu and Al, for a nanoparticle aspect ratio of 2.5 (data for Ag are shown for reference). The dielectric data of all three materials were taken from Ref. 29. Figure 4(a) shows that the wavelength of maximum radiative decay rate enhancement shows similar trends for all three metals. The optimal wavelength for Al is smaller than for the other metals due to its higher plasma frequency. The kink in the curve for Al at the equivalent diameter of about 80 nm is caused by the fact that, at the corresponding free-space wavelength of 800 nm (1.5 eV), Al exhibits an interband transition which causes additional Ohmic losses.[32] At this wavelength, the imaginary part of the dielectric function of Al is 30 times higher than that of Ag. Figure 4(b) shows that for prolate Al or Cu nanoparticles, the quantum efficiency of a low-quantum-efficiency emitter can be enhanced from 1% to above 60%. These data indicate that Ohmic losses are thus not necessarily a detrimental loss factor for plasmon-enhanced luminescence. Interestingly, the use of Al enables plasmon-enhanced luminescence frequencies well into the ultra-violet. Besides, Al and Cu are better compatible with silicon technology than Ag and Au and may thus enable the engineering of improved light emitters based on Si.

## V. Anisotropy-induced spectral separation of plasmon modes

Shape anisotropy can not only improve the coupling between an optical emitter and the dipole plasmon mode of a metal nanoparticle, as was discussed in Sec. IV. It also reduces the influence of dark higher-order plasmon modes on the luminescence properties of an emitter. In order to illustrate this effect, Figure 5 shows decay rate modifications as function of emission wavelength for an emitter in close proximity to a spherical metal nanoparticle. Subsequently, Figure 6 shows how these spectral characteristics change when the nanoparticle is made prolate-shaped.

Figure 5 shows the influence of a 60-nm-diameter Ag sphere on the decay rates of an optical emitter as a function of emission wavelength. The emitter is positioned at a fixed, relatively small distance of 3 nm from the sphere surface, in order to demonstrate the effect of dark plasmon modes on the decay rate of the emitter. The orientation of the source dipole was taken to be radial relative to the sphere. The refractive index of the embedding medium was set to 1.5, and the optical data for Ag were obtained from Ref. 29. Figure 5(a) shows the radiative and the total decay rates ($\Gamma_R$ and $\Gamma_{TOT}$, respectively) as a function of emission wavelength. Both $\Gamma_R$ and $\Gamma_{TOT}$ are normalized to the radiative decay rate of the emitter in the absence of the sphere $\Gamma_R^{REF}$. The total decay rate was obtained by taking into account coupling to all plasmon modes up to $l = 100$. Figure 5(a) shows that the radiative decay rate exhibits a variation of nearly three orders of magnitude; it peaks near 500 nm. The total decay rate, which includes both radiative decay and nonradiative decay that is associated with dissipation in the metal, peaks around 370 nm. The origin of this difference becomes clear from Fig. 5(b), which shows the contributions to the total decay rate of three plasmon modes with different angular mode number $l$: $l = 1$ (dipole mode), $l = 2$ (quadrupole mode), and $l = 30$. The dipole-mode contribution to the total decay rate peaks at the same wavelength as the radiative decay rate in Fig. 5(a), as expected, while the higher-order plasmon modes are resonant at shorter wavelengths.[14] The integrated effect of all higher-order modes on the total decay rate explains the different wavelengths of maximum radiative and total decay rate enhancement shown in Fig. 5(a).

Figure 5(c) shows the quantum efficiency $\eta = \Gamma_R / \Gamma_{TOT}$ of the emitter as a function of emission wavelength for four different reference quantum efficiencies $\eta^{REF}$ (i.e., quantum efficiency in the absence of the sphere): 0.1%, 1%, 10%, and 100%. The curves were obtained from the data shown in Fig. 5(a) by adding appropriate offsets to the nonradiative decay rate. The quantum efficiencies are close to zero at wavelengths below 400 nm, which is attributed to the strong excitation of dark higher-order plasmon modes and to the low radiative decay rate at these wavelengths. For $\eta^{REF} = 0.1\%$, the quantum efficiency is enhanced to about 6% at a wavelength of 500 nm. The quantum efficiency enhancement is thus as high as a factor 60. For $\eta^{REF} = 1\%$ and $\eta^{REF} = 10\%$, the maximum quantum efficiencies are about 21% and 28%, corresponding to enhancements of a factor 21 and 2.8, respectively. For $\eta^{REF} = 100\%$, the quantum efficiency spectrum hardly deviates from the spectrum for $\eta^{REF} = 10\%$. This resemblance is attributed to the fact that in both cases the excited-state decay is dominated by the excitation of plasmons. The quantum efficiency of the emitter is then equal to the outcoupling efficiency of plasmons to radiation, which is independent of $\eta^{REF}$. The quantum efficiency for $\eta^{REF} = 100\%$ is substantially reduced at a wavelength of 500 nm (by a factor 3.5), despite the fact that the radiative decay rate is enhanced by a factor 80.

For the spherical particles described here, the maximum quantum efficiency enhancement that can be achieved is limited by the contribution of dark higher-order modes. As we will show below, the key advantage of using prolate particles is that the effect of higher-order modes is suppressed. Figure 6 shows similar spectra as Fig. 5, but for a prolate nanoparticle with an aspect ratio of 2.5. The prolate volume is identical to the volume of the 60-nm-diameter Ag sphere considered in Fig. 5. The emitter is positioned along the major axis at a distance of 3 nm from the prolate surface, with the dipole orientation parallel to the axis. Figure 6(a) shows the radiative and total decay rates. A clear difference with the curves for the spherical particle in Fig. 5(a) is that the maximum of the radiative decay rate is shifted from 500 nm to 800 nm. At the same time the main peak in the total decay rate enhancement is hardly shifted. The origin of this difference becomes clear from the plots of the total decay rate contributions associated with different plasmon modes shown in Fig. 6(b). The dipole mode contribution to the total decay rate has strongly redshifted to 800 nm, while the higher-order contributions exhibit only a minor spectral shift. The latter is attributed to the fact that higher-order plasmon modes, with large numbers of closely spaced poles in the electric field, are less dependent on the surface curvature of the nanoparticle. Note that the dipole mode perpendicular to the long axis, which is blueshifted compared to the dipole mode of a sphere with identical volume, is not excited in the present configuration because of symmetry restrictions.

Figure 6(c) shows quantum efficiency spectra for $\eta^{\text{REF}} = 0.1\%$, 1%, and 10%. The spectrum for $\eta^{\text{REF}} = 100\%$ (not shown) coincides almost exactly with the spectrum for $\eta^{\text{REF}} = 10\%$. The maximum quantum efficiencies for $\eta^{\text{REF}} = 0.1\%$, 1%, and 10% are:

42%, 65%, and 69%, respectively. These quantum efficiencies are substantially higher than the values obtained for the spherical particle. For $\eta^{REF} = 0.1\%$ for example, the quantum efficiency enhancement increases from a factor 60 for the spherical particle to a factor 420 for the prolate nanoparticle. Clearly the spectral separation of dipolar and dark higher-order modes is a key factor that allows to obtain higher quantum efficiency enhancements near prolate particles.

The increased spectral separation of the plasmon modes in prolate metal nanoparticles is further illustrated in Fig. 7, which shows the photon energy $E_{max}$ at which the total decay rate associated with coupling to a particular plasmon mode is highest, plotted against the angular mode number $l$ of the plasmon mode. These photon energies correspond to the maxima shown in Figs. 5(b) and 6(b). The associated free-space wavelength $\lambda_{max}$ is indicated on the right-hand scale. For high angular mode numbers, both curves converge to the surface plasmon resonance energy of a flat interface between Ag and a dielectric with a refractive index of 1.5. This behavior, which readily follows from the expressions for the resonance condition for higher-order plasmon modes (see Eqs. A.1 and A.3), confirms that these modes, with closely-spaced charge nodes, depend less on the surface curvature than the lowest-order modes. The redshift for lower angular mode numbers is stronger for the prolate than for the spherical nanoparticle, again reflecting the increased spectral separation between radiative dipolar modes and dark higher-order modes for prolate particles. We note that the effect of this spectral separation is especially relevant at small emitter-nanoparticle separations (< 5 nm) due to the strong spatial decay of higher-order modes (see Eqs. A.1 and A.3, and Fig. 1(c)).

## VI. Conclusions

Using an analytical method developed by Gersten and Nitzan, modified to include radiative reaction and dynamic depolarization, we have provided insight in the physical processes that determine the coupling between prolate metal nano-antennas and optical emitters. We have shown that the luminescence quantum efficiency enhancement of an optical emitter placed close to a prolate metal nanoparticle is much higher than near a spherical particle. We have demonstrated that this effect is due to the geometry-induced electromagnetic field enhancement near the prolate tip, which enlarges the coupling between emitter and radiative dipolar plasmon modes (Fig. 3(a)). In particular for large nanoparticles (eq. diameter > 100 nm), that have near-unity radiation efficiency of the dipolar plasmon mode (Fig. 3(b)) this leads to a large radiative rate enhancement and thus strong quantum efficiency enhancement (Figs. 2(b)). The small influence of Ohmic losses on the plasmon outcoupling efficiency of large nanoparticles implies that metals that exhibit higher Ohmic losses than Ag and Au, such as Al and Cu, are suitable materials for optical nano-antennas, in particular when they are anisotropically shaped (Fig. 4). A second factor that determines the strong quantum efficiency enhancement for prolate metal nanoparticles is the spectral separation between the radiative dipolar and the dark higher-order modes. For increasing particle anisotropy the dipolar mode parallel to the long axis progressively shifts into the infrared, while the higher-order modes show much smaller shifts (Figs. 5(b), 6(b)). As a result, high emission quantum efficiencies are obtained at the redshifted dipolar plasmon wavelength. The physical insights obtained in this paper, illustrated by quantum efficiency improvements, are important for the optimization of emitted power as well.

**Acknowledgements**

Femius Koenderink is gratefully acknowledged for discussions on the topic of the paper. This work is part of the research program of the "Stichting voor Fundamenteel Onderzoek der Materie (FOM)", which is financially supported by the "Nederlandse organisatie voor Wetenschappelijk Onderzoek (NWO)". This work was also partially supported by NANONED, a nanotechnology program of the Dutch Ministry of Economic Affairs, and by AFOSR MURI Award No. FA9550-05-1-0450.

**Appendix: Expressions for the decay rates of an atom in the presence of a sphere according to the improved Gersten and Nitzan model**

This appendix lists the expressions for the total decay rate $\Gamma_{TOT}$ and the radiative decay rate $\Gamma_R$ of an excited atom in close proximity to a metal sphere, according to the improved Gersten and Nitzan (GN) model. The expressions for the radiative decay rates were obtained by implementing the correction factor for radiative reaction and dynamic depolarization in the original GN model,[16,17] The expressions for the *total decay rate* were obtained from the original Gersten and Nitzan model by implementing the same correction factor in the formula for the *nonradiative decay rate*. The fact that the correction factor accounts for radiation damping explains the transformation from nonradiative to total decay rate.

The atom, which is modeled as a classical dipole with dipole moment $\mu$, is positioned at a distance $d$ from the surface of a sphere with radius $a$ and dielectric constant $\varepsilon = \varepsilon' + i\varepsilon''$ in a homogeneous, non-absorbing medium with dielectric constant $\varepsilon_m$. We consider two dipole orientations: radial and tangential. For the radial dipole orientation, i.e., perpendicular ($\perp$) to the sphere surface, we obtain:

$$\frac{\Gamma_{TOT}^{\perp}}{\Gamma_R^{REF}} = 1 + \frac{3}{4(ka)^3} \sum_l l(l+1) \mathrm{Im}\left\{ C_n \frac{\varepsilon - \varepsilon_m}{\varepsilon + \frac{l+1}{l}\varepsilon_m} \left(\frac{a}{a+d}\right)^{2l+4} \right\}, \qquad (A.1)$$

$$\frac{\Gamma_R^{\perp}}{\Gamma_R^{REF}} = \left| 1 + 2C_1 \frac{\varepsilon - \varepsilon_m}{\varepsilon + 2\varepsilon_m} \left(\frac{a}{a+d}\right)^3 \right|^2, \qquad (A.2)$$

and for the tangential dipole orientation, i.e., parallel (//) to the sphere surface:

$$\frac{\Gamma_{TOT}^{//}}{\Gamma_R^{REF}} = 1 + \frac{3}{2(ka)^3}\sum_l (l+1)^2 \, \text{Im}\left\{ C_l \frac{\varepsilon - \varepsilon_m}{\varepsilon + \frac{l+1}{l}\varepsilon_m}\left(\frac{a}{a+d}\right)^{2l+4}\right\}, \quad (A.3)$$

$$\frac{\Gamma_R^{//}}{\Gamma_R^{REF}} = \left|1 - C_1 \frac{\varepsilon - \varepsilon_m}{\varepsilon + 2\varepsilon_m}\left(\frac{a}{a+d}\right)^3\right|^2, \quad (A.4)$$

where $\Gamma_R^{REF}$ refers to the radiative decay rate of the dipole emitter located in the same embedding medium in the absence of the sphere, $k = \sqrt{\varepsilon_m}\,\omega/c$, $\omega$ is the optical frequency (in radians per second), $c$ is the speed of light in vacuum, and $l$ is the angular mode number. $C_1$ is the correction factor for radiative reaction and dynamic depolarization:

$$C_1 = \frac{1}{1 - \frac{ik^3\alpha}{6\pi} - \frac{k^2\alpha}{4\pi a^2}}, \quad (A.5)$$

where $\alpha$ is the quasistatic polarizability of the sphere, which is defined as:

$$\alpha = 4\pi a^3 \frac{\varepsilon - \varepsilon_m}{\varepsilon + 2\varepsilon_m}. \quad (A.6)$$

For $l \neq 1$, no corrections are implemented, and thus $C_l = 1$ for $l \neq 1$.

The above expressions for $\Gamma_{TOT}^{\perp}$ and $\Gamma_{TOT}^{//}$ (Eqs. A.1 and A.3) refer to the total decay rate of an emitter with a luminescence quantum efficiency in the absence of the sphere. Results for emitters with intrinsic nonradiative decay were obtained by adding appropriate offsets to the total decay rate.

Note also that the expressions for the nonradiative decay rate that were obtained from the original GN model have been rewritten compared to Refs. 16 and 17 using the identity:

$$\text{Im}\left\{\frac{1}{\varepsilon+\frac{l+1}{l}}\right\} = -\frac{l}{2l+1}\text{Im}\left\{\frac{\varepsilon-1}{\varepsilon+\frac{l+1}{l}}\right\}. \qquad (A.7)$$

This procedure facilitates the implementation of the correction factor for radiative reaction and dynamic depolarization, as the dielectric resonance factor of the quasistatic polarizability (i.e. factor between brackets on the right side of Eq. A.7 for $l = 1$) can now readily be modified by adding the correction factor.

The correction for radiative reaction and dynamic depolarization can be implemented in the GN model for spheroids by taking $C_1$ and $\alpha$ as tensors, and by adapting Eq. (A.6) for spheroids.[26,30] The original GN expressions for radiative and total decay rate modifications near a spheroid can be found in Refs. 16 and 17 for a source position along the unique axis of a spheroid, and in Ref. 18 for an arbitrary source position in the vicinity of a spheroid.[33] The GN expressions for spheroidal nanoparticles rely on associated Legendre functions of the first and second kind.[34,35]

In some of the graphs, we have decomposed the radiative and the total decay rates in contributions that are associated with different plasmon modes. These contributions are calculated from Eqs. (A.1) to (A.4) by considering only one angular mode number $l$. In addition, the unity free-space contribution is subtracted. This approach facilitates the visualization of the distinct influence of different plasmon modes on the emission characteristics of the emitter. The decomposition in different mode contributions also uncovers a deficiency of the improved GN model: the fact that the system of source dipole and induced dipole are approximated by a single effective dipole, and interference effects between them are thus neglected, can result in a radiative decay rate associated with coupling to the dipole plasmon mode that is higher than the total decay rate associated with coupling to the same mode (see Fig. 3(b)). This deficiency of the GN model does not affect our conclusions. The reason is that interference between both dipoles is most prominent for configurations for which they have similar magnitude, whereas the induced dipole moment is much larger than the source dipole moment for configurations for which the decay rates are strongly enhanced.

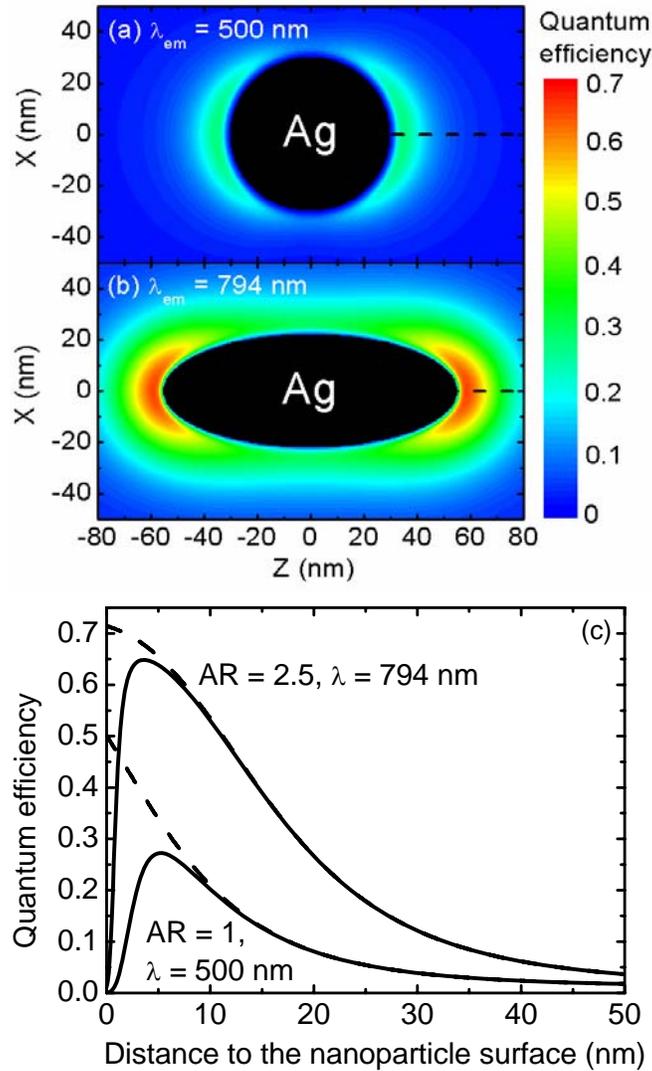

Figure 1. (color online) (a) and (b) Luminescence quantum efficiency of a dipole emitter as a function of emitter position in a plane through the particle's major axis for (a) a 60-nm-diameter Ag sphere, and (b) a Ag prolate with an aspect ratio of 2.5 and a volume that is equal to the volume of a 60-nm-diameter sphere. The orientation of the source dipole is taken parallel to the z-axis, and the refractive index of the embedding medium is 1.5. The emission wavelengths are the wavelengths of maximum radiative decay rate enhancement: 500 nm for the sphere and 794 nm for the prolate. The quantum efficiency

of the emitter in the absence of the sphere is 1%. The calculations were performed using the improved GN model by taking into account coupling to all modes up to $l = 100$. (c) Line traces of the luminescence quantum efficiency along the dashed lines indicated in (a) and (b), taking into account coupling to either all multipole modes up to $l = 100$ (solid lines), or to the dipole mode only (dashed lines).

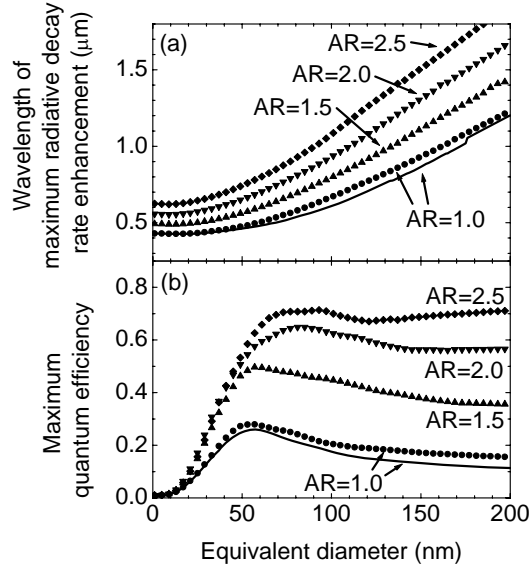

Figure 2. (a) Wavelength of maximum radiative decay rate enhancement associated with coupling to the dipole plasmon mode versus equivalent nanoparticle diameter, for a dipole emitter, the orientation of which is parallel to the major axis, positioned on the particle's major axis in close proximity to a prolate Ag nanoparticle, for four different nanoparticle aspect ratios (AR). (b) Maximum quantum efficiency versus equivalent diameter at the wavelength shown in (a), for the same four aspect ratios. The maximum quantum efficiency was found by varying the emitter-nanoparticle separation (see text). The refractive index of the embedding medium is 1.5, and the quantum efficiency of the emitter in the absence of the sphere is 1%. In both (a) and (b), the solid line for aspect ratio 1 was calculated based on exact electrodynamical theory; the lines with symbols were calculated with the improved GN model.

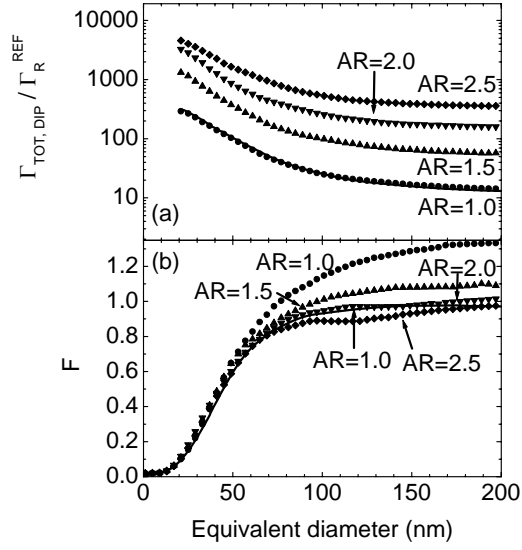

Figure 3. (a) Excited state decay rate associated with coupling to the dipole plasmon mode of the Ag nanoparticle $\Gamma_{\text{TOT,DIP}}$, normalized to the radiative decay rate in the absence of the sphere $\Gamma_R^{\text{REF}}$, versus equivalent diameter, plotted for different aspect ratios. (b) Fraction of the energy coupled to the dipole plasmon mode that is reradiated $F$. The parameters are the same as for Fig. 2, and the emission wavelength corresponds to the wavelength shown in Fig. 2(a). In both (a) and (b), the solid line for AR=1 was calculated based on exact electrodynamical theory; the lines with symbols were calculated with the improved GN model.

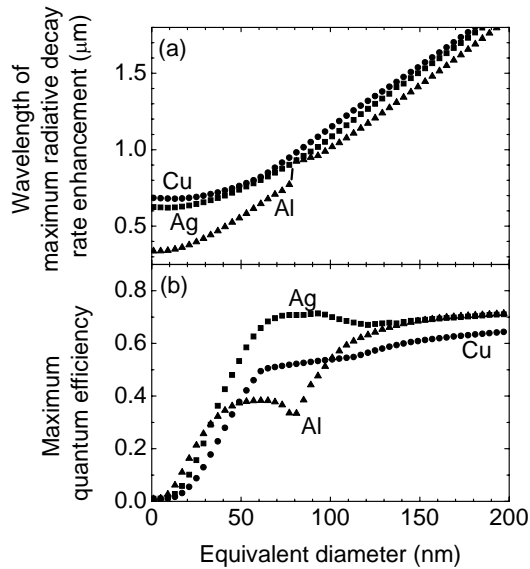

Figure 4. (a) Wavelength of maximum radiative decay rate enhancement associated with coupling to the dipole plasmon mode versus equivalent nanoparticle diameter, for a dipole emitter, the orientation of which is parallel to the major axis, positioned in close proximity to a prolate nanoparticle with an aspect ratio of 2.5, for three different nanoparticle materials: Ag, Cu and Al. (b) Maximum quantum efficiency versus equivalent diameter at the wavelength shown in (a), for the same three materials. The maximum quantum efficiency was found by optimizing the emitter-nanoparticle separation (see text). The refractive index of the embedding medium is 1.5, and the quantum efficiency of the emitter in the absence of the sphere is 1%. Calculation method: improved GN model.

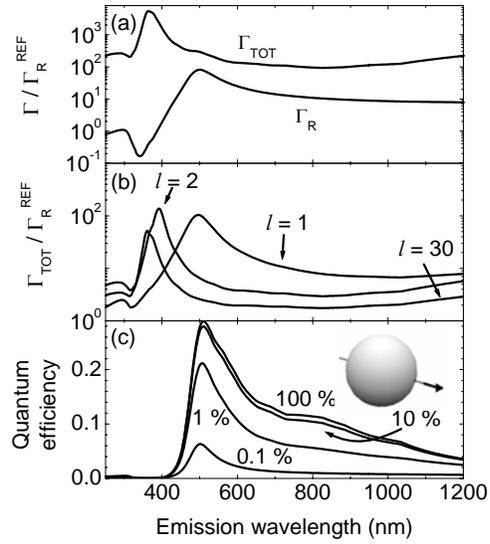

Figure 5. Calculations for a dipole emitter positioned at 3-nm distance from the surface of a 60-nm-diameter Ag sphere, embedded in a dielectric with a refractive index of 1.5. The dipole orientation is radial relative to the sphere. (a) Radiative and total decay rate enhancements as a function of emission wavelength. (b) Decay rate associated with coupling to plasmon modes with different angular mode numbers $l$. (c) Luminescence quantum efficiency of the emitter, for four different quantum efficiencies in the absence of the sphere: 0.1%, 1%, 10%, and 100%. Inset schematic representation of the Ag nanoparticle and the source dipole orientation. Calculation method: improved GN model.

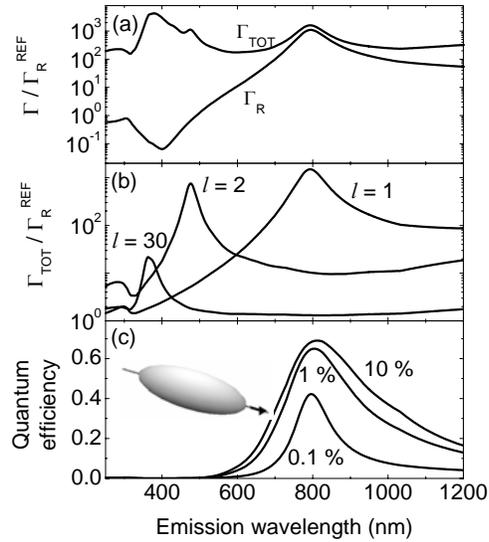

Figure 6. Calculations for a dipole emitter positioned on the particle's major axis at 3-nm distance from the tip of an Ag prolate with an aspect ratio of 2.5 and a volume equal to the volume of a 60-nm-diameter Ag sphere, embedded in a dielectric with a refractive index of 1.5. (a) Radiative and total decay rate enhancements as a function of emission wavelength. (b) Contributions to the total decay rate shown in (a) that are associated with coupling to plasmon modes with different angular mode numbers $l$. (c) Luminescence quantum efficiency of the emitter, assuming three different quantum efficiencies in the absence of the sphere: 0.1%, 1%, and 10%. Inset schematic representation of the Ag nanoparticle and the source dipole orientation. Calculation method: improved GN model.

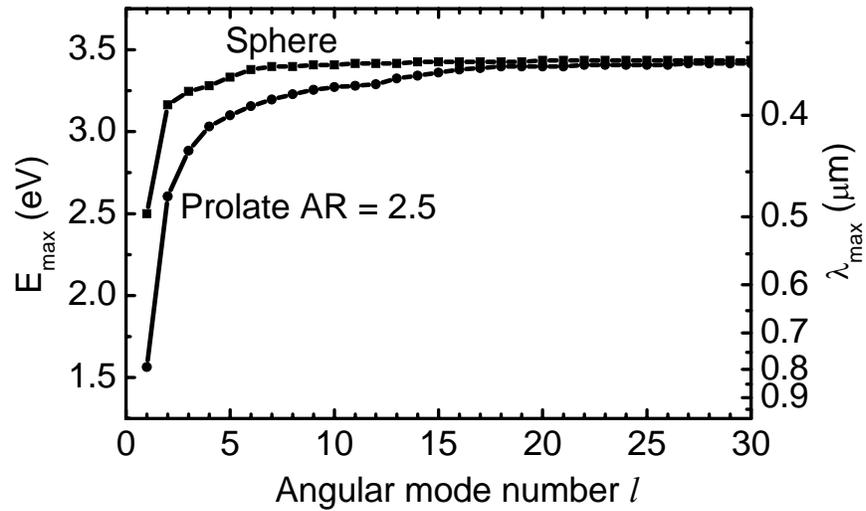

Figure 7. Photon energy $E_{max}$ at which the contribution to the total decay rate associated with coupling to a particular plasmon mode peaks versus angular mode number $l$ of the plasmon mode. The two emitter-particle configurations are the same as in Figs. 5 and 6. $\lambda_{max}$ is the free-space wavelength that corresponds to $E_{max}$. The lines are guides to the eye. Calculation method: improved GN model.